\documentstyle[prl,aps,epsfig,preprint]{revtex}
\newcommand{\be}{\begin{equation}}
\newcommand{\ee}{\end{equation}}
\newcommand{\bear}{\begin{eqnarray}}
\newcommand{\eear}{\end{eqnarray}}
\newcommand{\kone}{k_1}

\def\ktwo{k_2}

\def\dfp{\frac{d^4 p}{(2\pi)^4}}
\begin{document}
\draft
\title{Chiral vacuum alignment in  dense QCD}
\author{Taekoon Lee}
\address{Center for Theoretical Physics, Seoul National University, Seoul
151-742, Korea }
\maketitle
\begin{abstract}
We discuss an interesting possibility of nontrivial, quark-mass induced
chiral vacuum alignment in color-flavor locking phase of cold, dense QCD.
With the simplifying assumption that the gaps for quarks are identical to
those of antiquarks and  the light quark masses are given by
$m_u = m_d$ and $ m_s/m_d=15$, we find the true chiral vacuum can align only
to one of the discrete number of  directions in the continuum of chiral
vacua.  The alignment depends on the size of the diquark condensates, and
the vacuum transitions  between the discrete vacua caused by the
evolution of the diquark condensates
 can be  first order phase transition with vanishing or
nonvanishing latent heat,  depending on the vacua involved.
It is also shown that as $\mu \to \infty$, where $\mu$  is the
baryon chemical potential, parity is spontaneously broken through the
vacuum alignment.
\end{abstract}

The chiral symmetry $\text{SU}_L(3)\times \text{SU}_R(3)$ of quantum
chromodynamics (QCD) with three massless
quark flavors is spontaneously broken to $\text{SU}_{L+R}(3)$ at low energies.
Any point in the continuum of the chiral vacua, the coset space 
$[\text{SU}_L(3)
\times \text{SU}_R(3)]/\text{SU}_{L+R}(3)$, can be chosen as the vacuum, 
since any two vacua
are equivalent as far as physics is concerned.
However when quarks receive small (current) masses  the chiral symmetry
is then explicitly broken, and the degeneracy of the vacua is lifted.
The quark mass term then
picks up the lowest energy state in the continuum of the vacua as the
true chiral vacuum. This is called Dashen's chiral vacuum alignment 
\cite{dashen}.
To determine the true vacuum one has to find the quark-mass induced potential
that lifts the degeneracy of the vacua. Since the Nambu-Goldstone bosons
of the chiral symmetry breaking are small fluctuations about a point
in the continuum of  the vacua, the potential is nothing but the
meson mass term in the chiral Lagrangian of the nonlinearly
realized Nambu-Goldstone bosons. Therefore, the true vacuum is the
one that minimizes the potential
\bear
V(\Sigma)= -{\cal L}_m(\Sigma) \propto -{\mbox{Re}}[ {\mbox{Tr}}(m \Sigma)],
\label{meson_mass}
\eear
where $\Sigma \in \text{SU(3)}$ denotes the chiral fields
for the Nambu-Goldstone bosons, while  ${\cal L}_m(\Sigma)$
is the quark-mass induced meson mass term in the chiral Lagrangian.

When the true vacuum is nontrivial, that is, $\Sigma_0 \neq I$,
where $\Sigma_0$ is the vacuum that minimizes the potential,
some interesting phenomena could arise. For instance, if the quark mass matrix
were given as $m \propto {\mbox{Diag}}(-1,-1,-\delta)$ with $\delta > 1/2$,
then the vacuum $\Sigma_0$
would have an imaginary part in its matrix elements \cite{dashen}. Then
writing
$\Sigma=\Sigma_0 \exp(i \Pi^A T^A)$, where $\Pi^A$ and
$T^A$ denote the octet pseudo scalar mesons and the SU(3) generators,
respectively, it can be easily seen that the  mass term
(\ref{meson_mass}) could give rise to a spontaneous CP violation,
which enables, for example, $\eta$ decay into two pions. 
Of course, with the quark mass matrix realized in nature, 
the true vacuum is at $\Sigma_0 = I$,
and there is no such CP violation.

Recently, cold, dense quark system with large baryon chemical potential
$\mu$ has received strong interest \cite{arw,wilczek}. 
As well known, the 
Fermi surface
of such a dense system is unstable against Cooper pairing when an attractive 
force is present. 
At large  $\mu$ a dressed gluon exchange (in hard dense loop approximation) 
provides such a force, and causes the system to be in 
color-flavor locking phase in which quarks condensate in a pattern \cite{arw}
\bear
\langle \chi_i^a  \chi_j^b\rangle &\propto& 
k_1\, {(U_0)^a}_i {(U_0)^b}_j + k_2 \,{(U_0)^a}_j {(U_0)^b}_i \nonumber \\
\langle \bar{\varphi}_i^a\bar{\varphi}_j^b
\rangle &\propto& -[k_1\, {(V_0)^a}_i{(V_0)^b}_j +
k_2 \,{(V_0)^a}_j{(V_0)^b}_i],
\label{e1}
\eear
where $\chi^a_i,\varphi^a_i$, with $a$ the color index, $i=1,\cdots,3$,
the flavor index,
denote the two-component Weyl fermions for the left-handed quarks and the
complex conjugate of the right-handed quarks, respectively.
The unitary matrices $U_0, V_0 \in \text{U(3)}$ can be arbitrary in the
absence of axial $\text{U}(1)$ anomaly, but the anomaly effect, even though
small because of the suppression of  instanton effects at high density
\cite{pr37,rssv}, chooses a parity even vacuum \cite{rssv,ehhs} in which
\be
U_0=V_0.
\label{vac}
\ee

Upon the  condensation of the quarks the  symmetry of dense, massless QCD,
$\text{SU}_L(3)\times \text{SU}_R(3) \times  \text{U}_A(1) \times
\text{U}_B(1)$, where $\text{U}_A(1)$ and $\text{U}_B(1)$ denote the
axial and the baryon number symmetry, respectively, is spontaneously
broken to $\text{SU}_{L+R}(3)$, generating 10 Nambu-Goldstone bosons (mesons).
The $\text{U}_A(1)$ is not an exact symmetry, but at large $\mu$,
due to the suppression of anomaly, 
can be regarded  a good {\it approximate} symmetry.
As in vacuum QCD  a continuum of chiral vacua 
$[\text{SU}_L(3)\times \text{SU}_R(3) 
\times \text{U}_A(1)\times 
\text{U}_B(1)]/\text{SU}_{L+R}(3)$ arises upon the spontaneous
symmetry breaking.
Note, however, that the vacua connected by the $\text{U}_A(1)$
rotation are only approximately degenerate, because the  $\text{U}_A(1)$
is not an exact symmetry. It turns out that the vacuum of lowest energy
is parity even, hence the relation (\ref{vac}).

Now, when the current quark masses are turned on, the degeneracy of the chiral
vacua is lifted. The true vacuum can be picked up, as in vacuum QCD,
by minimizing the quark mass induced potential 
$V(\Sigma)=-{\cal L}_m(\Sigma)$ of dense QCD. 

The meson mass term ${\cal L}_m(\Sigma)$ in color-flavor locking phase has
a different form than in vacuum QCD because of the absence of left-right
quark condensates. For small quark masses,
${\cal L}_m(\Sigma)$ can be expanded in powers of the quark mass matrix.
The absence of a left-right quark condensate  due to the
suppression of instanton effects 
renders the leading term to be quadratic in quark mass \cite{arw}.
The most general form for the leading ${\cal L}_m(\Sigma)$, consistent with the
chiral symmetry and the condensates (\ref{e1}), is 
given by \cite{hlm,ss,rswz,mt,bbs}
\bear
{\cal L}_m(\Sigma)&=& A \,\,[
{\mbox{Tr}}(m^t\Sigma)]^2 + B\,\, {\mbox{Tr}}[(m^t\Sigma)^2] +C\,\, 
{\mbox{Tr}}(m^t\Sigma){\mbox{Tr}}(m^*\Sigma^\dagger) + \text{H.c.}\,, 
\label{e2}
\eear
where $\Sigma \in \text{U}(3)$ denotes the chiral fields for the 9 mesons.
Note that the Nambu-Goldstone boson associated with the baryon 
number symmetry remains exactly massless, and thus does not appear 
in the  mass term.

The coefficients $A,B$ and $C$ can be determined either by matching
the vacuum energy of (\ref{e2}) with that computed in the microscopic
theory \cite{ss} or by integrating out quark fields in the effective
Lagrangian of quarks and the mesons \cite{hlm}. Here we take the latter 
approach. Using the global color-chiral-axial-baryon number  symmetry
of dense QCD and the 
pattern of the diquark condensates (\ref{e1}) one can write an
effective chiral Lagrangian for the quarks and the Nambu-Goldstone
bosons as
\bear
{\cal L}&=& i \bar{\chi}^a_i\bar{\sigma}^{\nu}\partial_\nu\chi^a_i +
\mu \bar{\chi}^a_i\bar{\sigma}^{0}\chi^a_i +  i
\bar{\varphi}^a_i\bar{\sigma}^{\nu}\partial_
\nu\varphi^a_i -\mu \bar{\varphi}^a_i\bar{\sigma}^{0}\varphi^a_i
- [m_{ij}\chi^a_i\varphi^a_j
+\text{H.c.}]
\nonumber \\
&&+\left[ \chi^a_i (\Delta_\chi^\dagger)^{ab}_{ ij}\chi^b_j +
\varphi^a_i(\Delta_\varphi)^{ab}_{ ij}\varphi^b_j +\text{H.c.}\right] 
+{\cal L}_{{\mbox{{\tiny NG}}}}(U, V),
\label{e3}
\eear
where
\bear
 (\Delta_\chi^\dagger)^{ab}_{ij}&=&k_1 U^{*a}_iU^{*b}_j +k_2
U^{*a}_jU^{*b}_i, \nonumber \\
(\Delta_\varphi)^{ab}_{ ij}&=&-[k_1\, V^{a}_i\,V^{b}_j \,\,+k_2\,
V^{a}_j\,V^b_i].
\label{gapform}
\eear
The unitary matrices $U$ and $V$ denote the nonlinearly
realized Nambu-Goldstone
bosons arising from the symmetry breaking $SU_c(3)\times SU_L(3) \times
U_{A+B}(1) \to SU_{c+L}(3)$  through the $\chi \chi$ condensation and
$SU_c(3)\times SU_R(3) \times
U_{B-A}(1) \to SU_{c+R}(3)$
through the $\bar\varphi \bar\varphi$ condensation, respectively.
The ${\cal L}_{{\mbox{{\tiny NG}}}}(U, V)$ is the usual chiral Lagrangian
for the Nambu-Goldstone bosons alone\cite{hrz,cg}.  We also note that with
(\ref{gapform}) it was assumed, for simplicity,
the gaps for antiquarks are identical to those for quarks \cite{hlm}.
It can be easily seen that  the most important interactions for our
calculation,  those between the quarks and the
Nambu-Goldstone bosons, satisfy the Goldberger-Treiman relation. 
This Lagrangian can be regarded as an effective Lagrangian before
 color is gauged for the quarks and the 18 Nambu-Goldstone
bosons $U,V$.
Upon gauging color
it can be seen that
8 out of the 18 Nambu-Goldstone bosons are eaten
by the gluons via Higgs mechanism and there remain 10 color-singlet mesons
as low energy excitations.

Integrating out the quark fields, which corresponds to the evaluation of
quark one-loop diagrams in the background of constant $U,V$ fields
with two Dirac mass insertions (see Fig. 1), we obtain the mass term
(\ref{e2}) in which $\Sigma$ is given by
\be
\Sigma=UV^{\dagger}
\label{uvdagger}
\ee
and
\bear
A&=& i\int\dfp\left\{ (\kone^2+k_2^2)I_{8-}(p)I_{8+}(p)+
\kone(\kone+\ktwo/3)[I_{8-}(p)I_{8+}(p) \right.\nonumber\\
&& \left. -I_{1+}(p)I_{8-}(p)] +\kone(\kone+\ktwo/3)[I_{8-}(p)I_{8+}(p)-
I_{1-}(p)I_{8+}(p)]
\right.\nonumber \\
&&\left. +(\kone+\ktwo/3)^2[I_{8-}(p)I_{8+}(p)
-I_{1-}(p)I_{8+}(p) -I_{1+}(p)I_{8-}(p)\right.\nonumber\\
&&\left.+I_{1-}(p)I_{1+}(p)]\right\},
\nonumber\\
B&=&i\int\dfp\left\{ 2\kone\ktwo I_{8-}(p)I_{8+}(p)+
\ktwo(\kone+\ktwo/3)[I_{8-}(p)I_{8+}(p)\right.\nonumber\\
&& \left. -I_{1+}(p)I_{8-}(p)] +\ktwo(\kone+\ktwo/3)[I_{8-}(p)I_{8+}(p)-
I_{1-}(p)I_{8+}(p)] \right\}, \nonumber \\
C&=& -i\frac{1}{9} \int\dfp [-(p_0-\mu)^2+|\vec{p}|^2]
\left\{I_{8-}(p)I_{8+}(p)
-I_{1-}(p)I_{8+}(p)\right.\nonumber \\
&& \left. 
-I_{1+}(p)I_{8-}(p)+I_{1-}(p)I_{1+}(p)\right\},
\label{e6}
\eear
where 
\bear
I_{1\mp}(p)=1/[ -p_0^2+(|\vec{p}|\mp \mu)^2 +m_{1}^2], \hspace{.5in}
I_{8\mp}(p)=1/[ -p_0^2+(|\vec{p}|\mp \mu)^2 +m_{8}^2].
\eear
Here $m_1$ and $m_8$  are the Majorana masses for the singlet and octet
quarks, respectively, under the unbroken $\text{SU}_{L+R}(3)$ and are given as
\bear
m_{1}^2= (3 \kone +\ktwo)^2, \,\,\, m_{8}^2=\ktwo^2.
\eear

\begin{figure}
\begin{center}
\epsfig{file=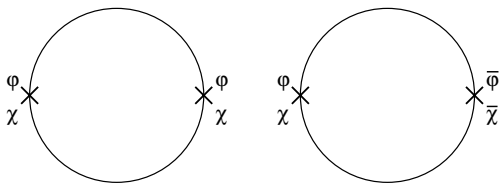, height=7.0cm, angle=90}
\end{center}
\caption{Diagrams to be evaluated in the constant background of
the chiral fields $U$ and $V$.}
\end{figure}

Generally the diquark condensates depend on energy, but here
we shall ignore
this fact and treat $k_i$ as constants.
Then the integration over the loop momentum can be
easily performed by doing contour integration over $p_0$ first
and then replacing  $d^3p \to 4\pi\mu^2 d |\vec{p}|$ for the integration over
the spatial components. It can be easily seen that $A$ and $B$
arise from the first diagram
in Fig.1 with $A, B \propto \Delta^2\ln
(\mu^2/\Delta^2)$, while $C$, coming from
the second diagram,  is given by
 $C\propto \Delta^4/\mu^2\ln
(\mu^2/\Delta^2)$, where $\Delta \sim k_i$.
Note that $C$ is suppressed by a factor
$\Delta^2/\mu^2$ compared to $A,B$.

To determine the true chiral vacuum at a given chemical potential
we have to minimize the potential $V(\Sigma)=-{\cal L}_m(\Sigma)$, with
${\cal L}_m(\Sigma)$ given by
(\ref{e2}) and (\ref{e6}). To demonstrate a nontrivial vacuum alignment, 
we shall now take a simplified form for the quark mass matrix, in which
the up and down quark masses are identical, as
\be
m = m_d\, \mbox{Diag} (1,1,\delta), \,\,\,
\delta \equiv m_s/m_d.
\label{e9}
\ee

With this quark mass matrix it is easy to see that the
$\Sigma$ that minimizes the potential must be of the form
\be
\Sigma = \mbox{Diag}(\alpha,\alpha,\beta) \,\,\,\text{with}\,\,\, 
|\alpha|^2=|\beta|^2=1,
\label{e10}
\ee
where $\alpha,\beta$ are complex variables.

Substituting (\ref{e9}) and (\ref{e10}) into (\ref{e2}) we obtain
\be
V= -2 m_d^2\left\{\mbox{Re}[ A(2 \alpha +\beta\delta)^2 +B(2 \alpha^2+
\beta^2\delta^2)
]+C \left|2 \alpha+\beta\delta\right|^2\right\},
\ee
which can also be written as
\be
V=-8 m_d^2 \left\{ (2 A+B) \cos^2\theta+(A+B)\delta^2/2\cos^2\phi
+A\delta\cos(\theta+\phi) +C\delta\cos(\theta-\phi)\right\}
\ee
by putting $\alpha=\exp(i\theta)$ and $\beta=\exp(i\phi)$.

We first notice that the potential is symmetric under the transformation
$(\alpha,\beta)\to (-\alpha,-\beta)$,  so the minima of the potential
must occur in
pairs. Secondly we observe that for arbitrary coefficients $A,B$, and $C$
the potential is stationary when $\theta, \phi$ satisfy 
\bear
\sin(\theta\pm\phi)=0, \,\,\,
\sin(2 \theta)=0,\,\,\,
\sin(2\phi)=0,
\eear
which have 8 common 
solutions corresponding to
the following $(\alpha,\beta)$ pairs
\be
(-i,-i), (-1,1), (1,1), (i,-i)
\label{e12}
\ee
and their  partners of opposite sign.
Of course none of these pairs needs necessarily minimize the potential,
but it can be shown numerically that the minima   occur always 
on one of these solutions.
This shows  that the true chiral vacuum can align only to one of these
eight directions.

In ideal situation we may know the dependence of the gap parameters $k_i$
on $\mu$, could choose the true chiral vacuum at a given chemical
potential from the above discrete vacua, and 
investigate vacuum transitions as the chemical
potential evolves. However, presently there is no reliable
calculation to determine the gaps as functions of the chemical potential
except when the chemical potential is
extremely large \cite{rs}, in which case Shwinger-Dyson
equation can be
used as an approximate gap equation and solved \cite{kpt,hms,sw}. 
Even in this case
the absolute
magnitudes of the gaps are yet to be determined, but, it may not be
so unreasonable to assume that $k_i/\mu$ are in the order of 0.1
at large chemical potential \cite{rssv,hms,ehhs}.

Taking into account this uncertainty  
we shall here treat  $k_i$ as free parameters and study
the vacuum transitions as $k_i$ vary. For definiteness, we shall put
$\delta=15$ and scan numerically the minima of the potential in the
$k_i$ parameter space defined by $0\le k_1/\mu\le 0.5$ and $ -0.5\le 
k_2/\mu\le 0.5$.
Note that $k_1$ can always be assumed positive, since its phase can be
rotated away by the baryon number symmetry of the Lagrangian (\ref{e3}).

The result of the numerical scanning is shown in Fig. 2.
As we see, the
parameter space is divided into four domains according to their vacuum
directions. In this figure the corresponding vacua of opposite sign
 to those in the group  (\ref{e12}) are not included.
The reason for this is that the numerical scanning shows there is
always a potential barrier between a vacuum in
the group (\ref{e12}) and the partners of those in the group  (\ref{e12}),
 so
in infinite volume limit the transitions between the vacua 
(\ref{e12}) and their  partners are negligibly small and can be ignored.
Therefore, only the transitions within the group of 
vacua in (\ref{e12}) need be considered.

\begin{figure}[b]
\begin{center}
\epsfig{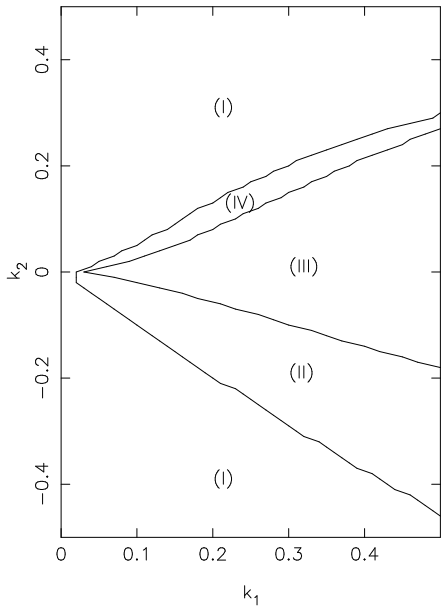}
\end{center}
\caption{Gap parameter space is divided into domains according to
their vacua. Each domain is associated with a unique vacuum. $k_i$ are in the
unit of $\mu=1$.}
\end{figure}

Because the vacuum can align only to a discrete number of directions the
transitions between vacua will be of first order phase transition.
To understand what happens at the transition, we look more carefully
at the potential on the boundaries between the domains.
For convenience, we shall designate the domains associated with the vacua
defined in (\ref{e12}) by (I),(II),(III), and (IV), respectively.

First, consider the transition  between the domain (I) and (II).
As one approaches the boundary from either side of (I) and (II)
it can be shown numerically that a potential valley opens up  in
$(\theta,\phi)$ space along the line $\phi=-\theta -\pi$,
with the bottom of the valley connecting the two points
$(-\pi/2,-\pi/2)$ and $(-\pi,0)$.
When exactly on the boundary, we have from
$V(\Sigma_{\mbox{\tiny I}})=V(\Sigma_{\mbox{\tiny II}})$, 
with $\Sigma_{\mbox{\tiny I,II}}$  defined
by (\ref{e10}) and  (\ref{e12}),
\be
(2 A +B) +(A+B)\delta^2/2-2 C \delta=0.
\ee
Then it is easy to see that on the bottom of the valley (i.e. along the
direction $\phi=-\theta -\pi$) the potential is
constant with $V=8m_d^2(A-C)\delta$. Thus, in this case there will be 
no latent heat released at the phase transition. 

Similarly, for the transition  between (II) and (III) it can be shown 
that a valley opens up along the line $\phi=0$
as  the boundary is approached, and that the relation 
\be
A+C=0,
\ee
holds on the boundary.
On the bottom of the valley
the potential is  given by
\be
V=-8m_d^2\left\{ (2 A+B) \cos^2\theta +  (A+B)\delta^2/2\right\},
\ee
which shows a barrier between the two vacua. This then indicates that 
there will be
latent heat released at the phase transition.  This may have some
important phenomenological consequence in dense systems.

In a similar fashion we can easily show that for the transition across the
domain (III) and (IV) a potential valley opens up along the direction
$\phi=-\theta$, and on the bottom of the valley the potential is constant with
$V=-8m_d^2(A-C)\delta$. In this case there will be no latent heat
released as in the transition between (I) and (II). Also for the transition 
between (I) and (IV) a valley opens up  along $\phi=-\pi/2$ direction and 
on the bottom of the valley the potential
is given by
\be
V=-4m_d^2(A+B)\delta^2 \cos^2\theta,
\ee
which shows a barrier between the two vacua, and consequently, nonzero latent
heat at the transition.

Could the phase transition of this kind occur when the
chemical potential increases from zero to an asymptotic value?
Although it is difficult to answer this question conclusively until we have
the quark-mass induced meson potential at an arbitrary value of $\mu$,
which at smaller $\mu$  would probably contain
the instanton-induced potential that gives mass to the $\eta'$ meson 
and
the old $\mbox{Tr}(m\Sigma)$ as well as  the quark-mass quadratic 
${\cal L}_m(\Sigma)$ in (\ref{e2}),
there is an interesting observation concerning this question.
It is well known that at large chemical potential the sextet components
of the condensates (\ref{e1}) is suppressed \cite{pr07,s09}, 
 thus $k_1\approx-k_2$,
which then suggests  the system must be in domain (I) at high density.
The vacuum associated with domain
(I) is $\Sigma_0=\mbox{Diag}(-i,-i,-i)$ that has an overall
factor $-i$ compared to the vacuum $\Sigma_0=I$
at zero density. Although this does not imply a  first order 
phase transition of the kind hitherto discussed, at least
it does suggest that the vacuum must shift from the unit 
matrix at zero density
to something else as the chemical potential increases.

This also leads to an interesting consequence of the vacuum alignment, namely
that parity must be
spontaneously broken at high density. The possibility of
parity violation in dense QCD was pointed out by Pisarski and Rischke
\cite{pr37}, and noted also in \cite{ehs,parity,arw}, who observed
that the axial U(1)
is almost a good symmetry at large $\mu$ due to the suppression of
instanton effects, so a vacuum with no parity symmetry can be as
easily formed as the parity even vacuum. Since the axial U(1) is never
exactly restored this possibility, however, cannot be realized 
with massless quarks. The lowest energy state is always
the parity even vacuum. However, when quarks are massive, the explicit
breaking
of the axial U(1) by quark masses can be more important
than the small anomalous breaking, so the true vacuum could break parity. 

Indeed, we see that this possibility is realized through the chiral
vacuum alignment.
Since the diquark condensates satisfy (\ref{vac}) in parity even vacuum
it can be seen using (\ref{uvdagger}) that only the
vacuum $\Sigma_0=I$ associated with the domain (III) is parity even.
Therefore, when the system is in domain (I), (II), and
(IV) parity is spontaneously broken. Moreover, since the system
should be in domain (I) as $\mu \to \infty$, parity must be
broken at high density.

To conclude, we have studied quark-mass induced  chiral vacuum alignment
in cold, dense QCD. When quarks are massless any point in the continuum of 
chiral vacua can be chosen as the vacuum, but when nonzero quark masses are
introduced, the true vacuum must be found via Dashen's procedure.
We have shown that in color-flavor locking phase at large chemical potential
the vacuum can align only to a discrete number of directions, and the nature
of the vacuum transitions is of first order with vanishing or nonvanishing
latent heat. 
It was  also shown that at high density parity is spontaneously 
broken through the  vacuum alignment. 
The consequence of
the first order phase transitions and  the parity violation  
may become important in dense systems like quark stars and 
heavy ion collisions.

Finally, we remark on the calculation of the meson masses.
Usually in meson mass calculation it was assumed that the vacuum is at
$\Sigma_0=I$, and the meson potential was expanded around this
vacuum to pick up the  meson spectrum. However, as we have seen, the vacuum
is not necessarily at the unit matrix and so this is not 
always correct. For correct meson masses one should first find  the true vacuum
and then expand the potential about it.

Useful comments by D.K. Hong are acknowledged.
This work was supported in part by the Korea Science and  Engineering
Foundation (KOSEF).




\end{document}